\documentclass[pof,preprint
]{revtex4-1}
\usepackage{graphicx}
\usepackage{amsmath}
\usepackage{amssymb}

\usepackage[utf8]{inputenc}
\usepackage[T1]{fontenc}
\usepackage{mathptmx}
\usepackage{etoolbox}
\usepackage{physics}

\usepackage{caption}
\usepackage[list=false]{subcaption}
\usepackage{flushend}

\usepackage{xcolor}
\usepackage{soul}

\usepackage{tikz}
\usepackage{tikz-3dplot}
\usepackage{adjustbox}
\usetikzlibrary{arrows}
\usetikzlibrary{calc, arrows.meta}

\draft 

\begin{document}

\title{Quantum Carleman Lattice Boltzmann Simulation of Fluids}
\author{Wael Itani}
\email{itani@nyu.edu}
\affiliation{Tandon School of Engineering, New York University, New York, NY 11201 , United States of America}

\author{Katepalli R. Sreenivasan}
\email{katepalli.sreenivasan@nyu.edu}
\affiliation{Tandon School of Engineering, New York University, New York, NY 11201 , United States of America}
\affiliation{Courant Institute of Mathematical Sciences, New York University, New York, NY 10012, United States of America}
\affiliation{Department of Physics, New York University, New York, NY 10003, United States of America}
\affiliation{Center for Space Science, New York University Abu Dhabi, Saadiyat Island, Abu Dhabi 129188, United Arab Emirates}

\author{Sauro Succi}
\thanks{Corresponding Author}
\email{sauro.succi@gmail.com}
\affiliation{Fondazione Istituto Italiano di Tecnologia,
Center for Life Nano-Neuroscience at la Sapienza, 00161 Roma, Italy}
\affiliation{Physics Department, Harvard University, Cambridge Massachusetts, USA}

\date{\today}

\begin{abstract}
We present a pedagogical introduction to a quantum computing algorithm for the simulation of classical fluids, based on the Carleman linearization of a second-quantized version of lattice kinetic theory. Prospects and limitations for the case of fluid turbulence are discussed and commented on.
\end{abstract}

\maketitle
\section{Introduction}

In 1982 Richard Feynman famously proclaimed that physics isn't
classical, hence it ought to be simulated on quantum computers \cite{feynmanSimulatingPhysicsComputers1982}.
It has ever since served as a source
of inspiration for quantum computing research.
Inspiration aside, the prospects of quantum computing are tantalizing
indeed, offering as they do, the potential chance of putting the quantum
superposition principle at use to explore and simulate in polynomial time problems that
present exponential barriers to classical algorithms \cite{groverQuantumMechanicsHelps1997}.
On general grounds, computer simulations
occupy the following four-quadrants of the physics-computing plane:
\begin{itemize}
\item[] CC: Classical computing for Classical physics;
\item[] CQ: Classical computing for Quantum physics;
\item[] QC: Quantum computing for Classical physics;
\item[] QQ: Quantum computing for Quantum physics.
\end{itemize}

To date, CC and CQ are by far the main and vastly more populated
quadrants: fluid dynamics and molecular dynamics belong in CC,
while electronic structure simulations are a prototypical CQ case. CC is often presented with polynomial computational complexity
(for instance, turbulence requires computing times that vary with the Reynolds number as its third or larger power)
and the CQ case often faces exponential complexity, a typical
case in point being the quantum many-body problem.

For nearly five decades Moore's law and parallel computing have 
successfully sustained an exponential growth of both CC and CQ 
quadrants, but for a few years now it is apparent that
the trend has slowed down, mostly on account of power consumption issues. In contrast, QQ offers, {\it in principle}, a natural escape from the
barrier of exponential complexity.

It is well known, however, that turning this potential into a concrete tool faces
daunting problems: for one, not all problems can be expressed
in terms of quantum computing algorithms; and, even when it is possible to do so,
their practical execution on actual quantum hardware
(QPU's) is confronted with efficiency issues and
the severe problem of decoherence. The so-called Quantum Advantage (QA), namely the expectation that
a quantum algorithm cannot be beaten by any classical one,
remains to be realized at present.
Not surprisingly, the most promising candidates are problems
in quantum chemistry and materials---namely, those centered
around the quantum many-body problem \cite{tacchinoQuantumComputersUniversal2020a}---although many other
applications (involving a search in high-dimensional spaces \cite{rogetGroverSearchNaturally2020}) hold promises
as well. In contrast, the QC quadrant remains largely unpopulated \cite{josephKoopmanvonNeumannApproach2020}.

This situation is hardly surprising because many classical problems
feature two major hurdles for quantum computing: 
nonlinearity and non-unitarity (dissipation). Nonlinearity implies dephasing \cite{hippertUniversalManybodyDiffusion2021}, hence loss of orthogonality,
because the rotation in the Hilbert space depends on the initial 
state vector: two different state vectors acted upon by the
same Hamiltonian rotate by different angles. 
Loss of orthogonality means loss of
information \cite{pokharelBetterthanclassicalGroverSearch2022a} and the time to tell apart two overlapping states scales
as $1/O$ where $O$ is the degree of orthogonality (overlap), with 
$O=0$ for the parallel case and $O=1$ for the orthogonal.
This is an inevitable problem at high Reynolds numbers, where loss of orthogonality
occurs very quickly.
Dissipation, on the other hand, cannot be dealt exactly by deterministic 
unitaries but typically requires a probabilistic implementation which comes with
a corresponding non-zero failure rate. Yet, given the paramount relevance of classical physics 
to science and engineering, an increasing group of quantum computing 
researchers is turning attention to this major challenge \cite{lopezDerivationMathematicalAnalysis2013,tennieQuantumComputersWeather2022}.

This paper occupies the QC quadrant, with specific focus
on the physics of fluids and, more precisely, the formulation of a quantum 
algorithm for fluids based on lattice kinetic theory. The motivations are straightforward: fluid turbulence faces a complexity of $Re^3$
or higher, $Re$ being the Reynolds number, basically the strength
of nonlinearity over viscous effects.
Most real life problems feature Reynolds numbers in the many-billions
(for an airplane $Re \sim 10^8$),  placing them well beyond the
reach of the best electronic supercomputers, now in
the exascale range.
In contrast, a $Q$ qubit computer (IBM is currently at Q=433, nearing
500 in next two years), offers $2^{Q} \sim 10^{3Q/10}$ binary
degrees of freedom. A turbulent flow at a given Reynolds number
contains $Re^{9/4}$ degrees of freedom (or more), hence the (minimum) 
number of qubits required to represent such a turbulent flow is given by
\begin{equation}
Q \sim \frac{15}{2} log(Re).
\end{equation} 
This shows that $Q=60$ is already matching the exascale capacity \cite{succiExascaleLatticeBoltzmann2019}, while the Reynolds number would skyrocket to about $Re \sim 10^{20}$ with
$Q=120$,  
far beyond the capacity of any foreseeable classical computer.
Hence, the potential definitely exists, although its actual realization
faces mounting difficulties with increasing Reynolds numbers.
 
The specific focus on lattice kinetic theory is motivated by the hope
that the substantial advantages offered by such formulation in the case
of classical fluids can somehow be transferred to the quantum realm.
In particular, we refer to the fact that nonlinearity and nonlocality are
disentangled, i.e., streaming is  nonlocal but linear, while collisions
are nonlinear but local. Since the first simulation based on the lattice Boltzmann model in 1989 \cite{higueraSimulatingFlowCircular1989}, the adoption of the model has grown, especially with the advent of GPUs which could parallelize the model's algorithms thanks to the aforementioned separation. Lattice models could, similarly, utilize the parallelism afforded by quantum computers \cite{bharadwajQuantumComputationFluid2020a}. Similar to the case with classical computers \cite{higueraSimulatingFlowCircular1989}, algorithms for the simulation of classical fluids on a quantum computer using lattice kinetic theory started with focus on lattice gas \cite{yepezLatticeGasQuantumComputation1998,yepezQuantumComputationFluid1999,vahalaQuantumLatticeGas2008} to lattice Boltzmann \cite{yepezOpenQuantumSystem2006,yepezEfficientQuantumAlgorithm2002,vahalaQuantumLatticeGas2008}. However, we make a distinction between those that attempt to achieve a quantum analogue \cite{mezzacapoQuantumSimulatorTransport2015} versus those which use of a quantum computer is due to the advantage it affors when carrying out the arithmetic \cite{moawadInvestigatingHardwareAcceleration2022a,steijlQuantumAlgorithmsFluid2020,steijlQuantumCircuitImplementation2023}

In contrast, the Navier-Stokes fluid self-advection
operator which is non-local and non-linear at once. The result is that information
travels along space-time material lines defined by the flow velocity
$\frac{dx}{dt}=u(x,t)$, while in lattice kinetic theory information moves along
straight lines defined by discrete velocities $v$, namely $\frac{dx}{dt}=v$,
where $v$ is a constant. In addition, thanks to the extra dimensions inherent to the phase space $\Gamma = (x,v)$, both the pressure and the strain-rate are locally available online with
no need for solving the Poisson equation requiring second order spatial derivatives. 
This is expected to simplify the structure of the Carleman linearization in comparison to the
fluid case, which involves complex pressure-velocity and pressure-strain-rate correlations. Furthermore, the Carleman-linearized formulation would benefit from quantum algorithms designed for a high-dimensional phase-space \cite{pfefferHybridQuantumclassicalReservoir2022a}.

\section{The Lattice Boltzmann Equation in the Mode-Coupling Form}
\begin{figure*}
\centering
\includegraphics[scale=1]{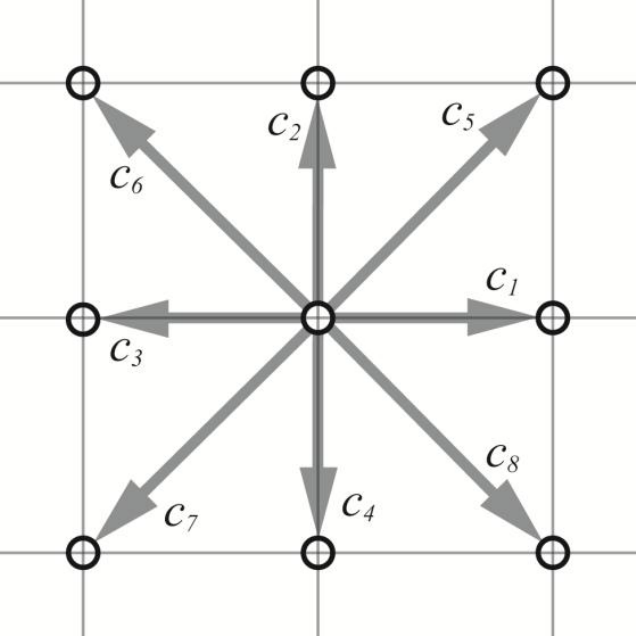}
\caption{A single site in the standard D2Q9 lattice\label{fig:d2q9fivefam}}
\end{figure*}
We consider the lattice Boltzmann (LB) equation
in single-time relaxation form and one spatial dimension, without loss
of generality:
\begin{widetext}
\begin{equation}
\label{LBC1}
f_i(x+c_i \Delta t,~t+\Delta t) -f_i(x,t)=(1-\omega)f_i(x,t)+\omega f_i^{eq}(x,t).
\end{equation}
\end{widetext}
Here $f_i(x,t)=f(x,v,t) \delta(v-c_i)$ is a set of discrete Boltzmann
distributions moving with discrete velocity  
$c_i, i=1,b$, chosen according to a suitable symmetry group; $\omega = \Omega \Delta t$.
The LHS is the free streaming along direction $c_i$ while the RHS stands for
collisional relaxation towards the local equilibrium $f_i^{eq}$ on a typical
timescale $\tau=1/\Omega$.
The local equilibrium compatible with the Navier-Stokes equation of 
incompressible fluid dynamics reads as \cite{succiLatticeBoltzmannEquation2001,succiLatticeBoltzmannEquation2018}:

\begin{equation}
\label{LEQ}
f_i^{eq} = w_i \rho(1+\frac{c_iu}{c_s^2}+ \frac{Q_iu^2}{2c_s^4}), 
\end{equation}
where $w_i$ is a suitable set of weights normalized to unity.
In the above, $\rho=\sum_i f_i$  is the fluid density, $u= \sum_i f_i c_i$ is the fluid 
current and $Q_i = c_i^2 -c_s^2$, $c_s^2$ being the lattice sound speed,
is a constant $O(1)$ in lattice units.

In $d=2$, the standard D2Q9 model, namely the second-rank tensor, can be obtained through taking the tensor product of the first-rank vector $(-1,0,1)$ for D1Q3.
Likewise, the $D3Q27$ lattice directions
are defined by the third-rank tensor product of D1Q3 = $(-1,0,1)$ (see Fig.2).

\begin{figure}[ht]
\centering
\captionsetup[subfigure]{justification=centering}

\subcaptionbox[D3Q27]{D3Q27\label{subfig:d3q27}}
[3cm]
{\resizebox{3cm}{!}{
\tdplotsetmaincoords{74}{113}
         \begin{tikzpicture}[tdplot_main_coords,
axis/.style={thick, ->, >=stealth'}]

\coordinate (O) at (0.5,0.5,0.5);

\def \a {1}       
\def \b {1}       
\def \c {1}       

 \foreach \u in {0,1,...,\a}
    \foreach \v in {0,1,...,\b}
      \foreach \w in {0,1,...,\c}
        \draw[very thin,gray] (\u,\v,0) -- (\u,\v,\w); 
 \foreach \u in {0,1,...,\a}
    \foreach \v in {0,1,...,\b}
      \foreach \w in {0,1,...,\c}
        \draw[very thin,gray] (\u,0,\w) -- (\u,\v,\w); 
 \foreach \u in {0,1,...,\a}
    \foreach \v in {0,1,...,\b}
      \foreach \w in {0,1,...,\c}
        \draw[very thin, gray] (0,\v,\w) -- (\u,\v,\w);

\draw plot [mark=*, mark size=1] coordinates{(O)};

 \foreach \u in {0,1,...,\a}
    \foreach \v in {0,1,...,\b}
      \foreach \w in {0,1,...,\c}
        \draw[thin,-latex,black](O) -- (\u,\v,\w);
        
    \foreach \v in {0,1,...,\b}
      \foreach \w in {0,1,...,\c}
        \draw[thin,-latex,black](O) -- (0.5,\v,\w);
 \foreach \u in {0,1,...,\a}
      \foreach \w in {0,1,...,\c}
        \draw[thin,-latex,black](O) -- (\u,0.5,\w);
 \foreach \u in {0,1,...,\a}
    \foreach \v in {0,1,...,\b}
        \draw[thin,-latex,black](O) -- (\u,\v,0.5);

 \foreach \u in {0,1,...,\a}
        \draw[thin,-latex,black](O) -- (\u,0.5,0.5);
    \foreach \v in {0,1,...,\b}
        \draw[thin,-latex,black](O) -- (0.5,\v,0.5); 
      \foreach \w in {0,1,...,\c}
        \draw[thin,-latex,black](O) -- (0.5,0.5,\w); 

\end{tikzpicture}}}
\hspace{0.1\textwidth}
\subcaptionbox[D2Q9]{D2Q9\label{subfig:d2q9}}
[3cm]
{\resizebox{3cm}{!}{
\tdplotsetmaincoords{74}{113}
         \begin{tikzpicture}[tdplot_main_coords,
axis/.style={thick, ->, >=stealth'}]

\coordinate (O) at (0.5,0.5,0.5);

\def \a {1}       
\def \b {1}       
\def \c {1}       

 \foreach \u in {0,1,...,\a}
    \foreach \v in {0,1,...,\b}
      \foreach \w in {0,1,...,\c}
        \draw[very thin,gray] (\u,\v,0) -- (\u,\v,\w); 
 \foreach \u in {0,1,...,\a}
    \foreach \v in {0,1,...,\b}
      \foreach \w in {0,1,...,\c}
        \draw[very thin,gray] (\u,0,\w) -- (\u,\v,\w); 
 \foreach \u in {0,1,...,\a}
    \foreach \v in {0,1,...,\b}
      \foreach \w in {0,1,...,\c}
        \draw[very thin, gray] (0,\v,\w) -- (\u,\v,\w);

\draw plot [mark=*, mark size=1] coordinates{(O)};

    \foreach \v in {0,1,...,\b}
      \foreach \w in {0,1,...,\c}
        \draw[thin,-latex,black](O) -- (0.5,\v,\w);

    \foreach \v in {0,1,...,\b}
        \draw[thin,-latex,black](O) -- (0.5,\v,0.5); 
      \foreach \w in {0,1,...,\c}
        \draw[thin,-latex,black](O) -- (0.5,0.5,\w); 

\end{tikzpicture}

\tdplotsetmaincoords{74}{113}
         \begin{tikzpicture}[tdplot_main_coords,
axis/.style={thick, ->, >=stealth'}]

\coordinate (O) at (0.5,0.5,0.5);

\def \a {1}       
\def \b {1}       
\def \c {1}       

 \foreach \v in {0,1,...,\b}
      \foreach \w in {0,1,...,\c}
        \draw[very thin,gray] (0.5,\v,0) -- (0.5,\v,\w); 
 \foreach \v in {0,1,...,\b}
      \foreach \w in {0,1,...,\c}
        \draw[very thin,gray] (0.5,0,\w) -- (0.5,\v,\w);

\draw plot [mark=*, mark size=1] coordinates{(O)};

    \foreach \v in {0,1,...,\b}
      \foreach \w in {0,1,...,\c}
        \draw[thin,-latex,black](O) -- (0.5,\v,\w);

    \foreach \v in {0,1,...,\b}
        \draw[thin,-latex,black](O) -- (0.5,\v,0.5); 
      \foreach \w in {0,1,...,\c}
        \draw[thin,-latex,black](O) -- (0.5,0.5,\w); 

\end{tikzpicture}}}
\hspace{0.1\textwidth}
\subcaptionbox[D1Q3]{D1Q3\label{subfig:d1q3}}
[3cm]
{\resizebox{3cm}{!}{
\tdplotsetmaincoords{74}{113}
         \begin{tikzpicture}[tdplot_main_coords,
axis/.style={thick, ->, >=stealth'}]

\coordinate (O) at (0.5,0.5,0.5);

\def \a {1}       
\def \b {1}       
\def \c {1}       

 \foreach \u in {0,1,...,\a}
    \foreach \v in {0,1,...,\b}
      \foreach \w in {0,1,...,\c}
        \draw[very thin,gray] (\u,\v,0) -- (\u,\v,\w); 
 \foreach \u in {0,1,...,\a}
    \foreach \v in {0,1,...,\b}
      \foreach \w in {0,1,...,\c}
        \draw[very thin,gray] (\u,0,\w) -- (\u,\v,\w); 
 \foreach \u in {0,1,...,\a}
    \foreach \v in {0,1,...,\b}
      \foreach \w in {0,1,...,\c}
        \draw[very thin, gray] (0,\v,\w) -- (\u,\v,\w);

\draw plot [mark=*, mark size=1] coordinates{(O)};

    \foreach \v in {0,1,...,\b}
        \draw[thin,-latex,black](O) -- (0.5,\v,0.5); 
\end{tikzpicture}

\tdplotsetmaincoords{74}{113}
         \begin{tikzpicture}[tdplot_main_coords,
axis/.style={thick, ->, >=stealth'}]

\coordinate (O) at (0.5,0.5,0.5);

\def \a {1}       
\def \b {1}       
\def \c {1}       

 \foreach \u in {0,1,...,\a}
    \foreach \v in {0,1,...,\b}
      \foreach \w in {0,1,...,\c}
        \draw[very thin,white] (\u,\v,0) -- (\u,\v,\w); 
 \foreach \u in {0,1,...,\a}
    \foreach \v in {0,1,...,\b}
      \foreach \w in {0,1,...,\c}
        \draw[very thin,white] (\u,0,\w) -- (\u,\v,\w); 
 \foreach \u in {0,1,...,\a}
    \foreach \v in {0,1,...,\b}
      \foreach \w in {0,1,...,\c}
        \draw[very thin,white] (0,\v,\w) -- (\u,\v,\w);

\draw plot [mark=*, mark size=1] coordinates{(O)};

    \foreach \v in {0,1,...,\b}
        \draw[thin,-latex,black](O) -- (0.5,\v,0.5); 
\end{tikzpicture}}}
\caption{Different lattice configurations in three, two and one dimensions}
\label{fig:lattices}
\end{figure}
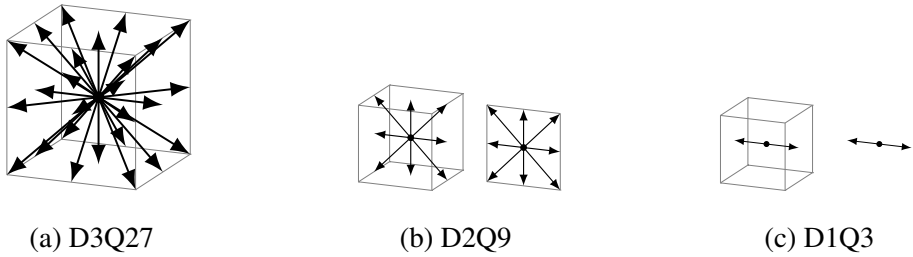


\subsection{Mode Coupling Form}

By the definitions of $\rho$ and $u$, the local equilibrium can be written 
in the mode-coupling form as
\begin{equation}
f_i^{eq} =  L_{ij}f_j+Q_{ijk} f_j f_k, 
\end{equation}
where  
\begin{eqnarray}
L_{ij}=w_i(1+c_i c_j/c_s^2)\\
Q_{ijk} = w_i Q_i c_j c_k/2c_s^4.
\end{eqnarray}
As a result the LBE takes the form
\begin{equation}
f_i(x+c_i, t+1) = A_{ij}f_j + B_{ijk}f_j f_k, 
\end{equation}
where we have set $A_{ij}=\delta_{ij}-\omega L_{ij}$ and $B_{ijk}=\omega Q_{ijk}$.
Note that the RHS has two fixed points, an unstable trivial vacuum $f_i=0$
and a non-trivial stable one, $f_i = f_i^{eq}$.
This is the desired mode-coupling form which proves expedient 
to the Carleman formulation, to be discussed next.
Before doing so, let us note that, owing to conservation laws, the above matrices
display these sum-rules: 
\begin{eqnarray}
\label{SUMRULES}
\sum_i w_i = 1,\;
\\\sum_i L_{ij}=\sum_j L_{ij}=1,\;
\\\sum_i Q_{ijk} = 0.
\end{eqnarray}

As we shall show, these conservation laws lay at the ground of the 
exact closure at the second order of the Carleman procedure
for the homogeneous kinetic equation.

\section{Carleman Linearization}

As is well known, the Carleman linearization transforms
nonlinear equation into an equivalent infinite-dimensional
linear system.
Let us illustrate the idea for the simple case of 
the logistic equation. 

\subsection{Carleman Treatment of the Logistic Equation}

Consider the logistic equation
\begin{equation}
\partial_t f = -af+bf^2;\;\;\;f(0)=f_0
\end{equation}
with $a,b$ both positive.
This can be seen as the homogeneous version $(\partial_x f =0)$ of a corresponding
kinetic equation with two fixed points, a stable one, $f=0$, and
an unstable one, $f=a/b=K$, where $K$ is the so-called carrying-capacity.
Note that $R=b/a=1/K$ measures the strength of the nonlinearity (and is akin to the Reynolds number).
The exact solution reads 
\begin{equation}
f(t) = \frac{f_0 e^{-at}}{1-\frac{f_0}{K}(1-e^{-at})}.
\end{equation}
For $0<f_0<K$ this decays to zero, while for $f_0>K$ it develops
a finite-time singularity in a time lapse $at_{sing} \sim K/f_0$.  

The Carleman procedure is readily shown to lead to the
infinite chain of ODE's
\begin{equation}
\frac{df_k}{dt} = - k(a f_k +  b f_{k+1}),\;\;\; k=1, k_{max}
\end{equation}
with initial conditions $f_k(0)= f_0^k$.
The first order truncation $f_2=0$ yields a pure exponential
decay $f_1^{(0)}(t)= f_0 e^{-at}$. Besides missing the slow
transient, this also provides an incorrect asymptotic amplitude.
However, as long as $f_0/K \ll 1$, it is readily shown
that a simple Euler time marching with sufficiently small time step 
$a \Delta t \ll 1$, delivers a pretty accurate solution
with just a few Carleman iterates.
However, the convergence deteriorates rapidly as $f_0/K \to 1$.

\begin{figure*}
\centering
\begin{subfigure}{\columnwidth}
\includegraphics[scale=0.2]{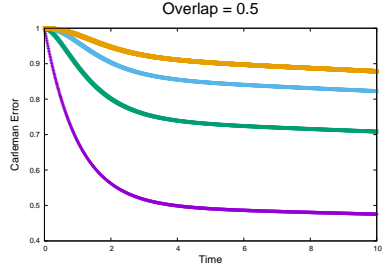}
\caption{$Rf_0 = 0.5$\label{subfig:error1}}
\end{subfigure}
\\
\begin{subfigure}{\columnwidth}
\includegraphics[scale=0.2, angle = 270]{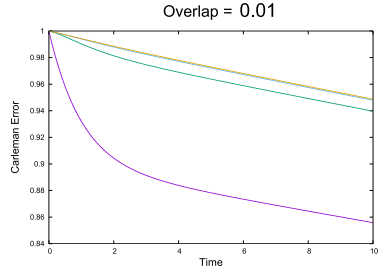}    
\caption{$Rf_0 = 0.01$\label{subfig:error2}}
\end{subfigure}
\\
\begin{subfigure}{\columnwidth}
\includegraphics[scale=0.2]{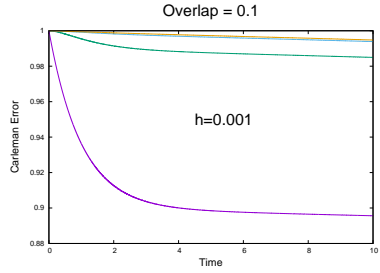}    
\caption{$Rf_0 = 0.1$\label{subfig:error3}}
\end{subfigure}

\caption{Absolute Carleman error with overlap $Rf_0=0.5$ (\ref{subfig:error1}) 
and $Rf_0=0.01$ (\ref{subfig:error2}), with time
step $\Delta t=0.01$, with truncation from the first (orange) to the fourth (violet) order.
(\ref{subfig:error3}) is the absolute Carleman error with overlap $Rf_0=0.1$ and time
step $\Delta t = 0.001$, with truncation from the first to the fourth order.
}
\end{figure*}
\subsection{Carleman Linearization for Kinetic Theory versus Fluid Dynamics} 

In the spirit of developing a Carleman-based quantum algorithm for fluids, it is natural
to focus the attention on the Navier-Stokes equations. 
A simple inspection shows that the Carleman linearization of the Navier-Stokes 
equations meets  with a number of complications due to correlations between the fluid flow
$u_\mu$, $\mu=x,y,z$, the strain sensor $D_{\mu \nu} = \partial_\mu u_\nu$ and the fluid
pressure $p$. In particular, a dynamic equation for the fluid pressure is required
to generate the pressure-velocity and pressure-strain correlators, and
the dynamic equations for the corresponding correlators are quite cumbersome.

On the other hand, the Carleman linearization of the kinetic equation gives rise to
a hierarchy of multiscalars, $f_i \to f_i f_j \to f_i f_j f_k \to ...$ and 
the corresponding dynamic equations remain first order in space and time, which
we expect to provide a significant advantage for the formulation of a
quantum computing algorithm.  
More importantly,  thanks to the basic mass-momentum conservation laws, there
is no need to track all the multiscalars above, but only a limited 
subset of linear combinations therefrom. 

Either way, the key question is the convergence as a function of the Reynolds number.
Based on the logistic results, one would expect Carleman convergence to occur 
under the constraint 
\begin{equation}
\label{CCO}
|\frac{f-f^{eq}}{f^{eq}}| \ll 1/Re.
\end{equation}
This looks quite demanding at high Reynolds number, although 
a moment's thought reveals that such a constraint is 
fully in line with the hydrodynamic limit of kinetic theory.

To this end, let us remind that hydrodynamics emerges from the kinetic theory
in the limit of weak departure from local equilibrium, or, differently 
restated, for small Knudsen numbers $Kn = \lambda/L \sim |f-f^{eq}|/f^{eq}\ll 1$, 
where $\lambda$ is the molecular mean free path and $L$ a characteristic hydrodynamic scale.
The next observation is that the Knudsen number scales inversely with
the Reynolds number $Re=UL/\nu$, according to the so-called von K\'arm\'n relation
\begin{equation}
Kn = \frac{Ma}{Re},
\end{equation}
where $Ma$ is the Mach number.
Taking $Ma \sim O(1)$, and $f^{eq} \sim 1$, the above relation coincides
with the constraint Eq.~(\ref{CCO}). 
For a typical car, we have $Re \sim 10^7$, indicating that the departure from
local equilibrium is on the order of the seventh digit.
This is unquestionably a stringent request,  mitigated however by the hydrodynamic
conservation laws, as we shall discuss shortly.

\subsection{Carleman Lattice Boltzmann (CLB) Scheme}

We will now discuss this last point by providing the explicit form of
the Carleman Lattice Boltzmann (CLB) scheme to first and second orders.

\subsubsection{First-order CLB}

To the first order the standard Lattice Boltzmann (LB) equation for $f_i$ is 
\begin{equation}
\label{LBC1firstorder}
f_i(x+c_i \Delta t, t+\Delta t) = (1-\omega) f_i(x,t) + \omega g_i(x,t),
\end{equation}
where $\omega = \Omega \Delta t$, as before, and we have set $g_i \equiv f_i^{eq}$ to 
denote a generic collisional attractor.
At this stage, the Carleman array of variables is just $F_1=[f_i]$,

It is to be noted further that the negative lattice viscosity, also known as propagation
viscosity $\nu_P = -\frac{1}{2}$ (in lattice units $\Delta x = \Delta t =1$),
can be incorporated within an effective physical viscosity, $\nu = c_s^2 (1/\omega-1/2)$,
thereby permitting one to march in large steps $\Delta t \sim O(\tau)$ without losing stability.  
This stands in contrast to Euler marching for ODEs, which requires $\Delta t \ll \tau$.

\subsubsection{Second-order CLB}

We write the LB equation at two different locations $x_i=x+c_i$ and $x_j=x+c_j$, as
\begin{equation}
\label{LBC121}
f_i(x_i,t+1) = (1-\omega) f_i(x,t) + \omega g_i(x,t)
\end{equation}
\begin{equation}
\label{LBC122}
f_j(x_j,t+1) = (1-\omega) f_j(x,t) + \omega g_j(x,t)
\end{equation}
where we have $\Delta t=1$ for convenience.

Multiplying one equation by the other, we obtain

\begin{widetext}
\begin{equation}
\label{}
f_{ij}(x_i,x_j,t+1) 
= (1-\omega)^2 f_{ij}(x,x,t)  
+ 2 \omega (1-\omega) h_{ij}(x,x;t)
+ \omega^2 g_{ij}(x,x,t),
\end{equation}
\end{widetext}
where we have set
\begin{equation}
g_{ij}= g_i g_j
\end{equation}
for the double-equilibrium and
\begin{equation}
h_{ij} = \frac{1}{2}(f_ig_j+g_if_j)
\end{equation}
for the semi-equilibrium.

Several comments are in order. First, we note that for $\omega=0$, we still have an exact free-streaming 
formulation $f_{ij}(x_i,x_j,t+1)= f_{ij}(x,x,t)$. Second, the second term on the right-hand-side takes the symbolic form
form $h=fLf+fQff$, while the third one gives $g=LfLf+LfQff+QffLf+QffQff$, indicating
coupling with third and fourth order Carleman variables.
Clearly, Carleman truncation at order two retains only $fLf$ and $LfLf$, although
a better approximation might be obtained by replacing $f$ in the higher order terms
with a zero-velocity equilibrium, $f_i \sim \rho w_i$.
We also observe that the above notation invites a natural analogy with tensor 
networks which might be worth exploring for the future \cite{gourianovQuantuminspiredApproachExploit2022}. 

Most important of all, as a consequence of streaming, the second order 
Carleman array involves a pair (Carleman pairs) of locations $x_i$
and $x_j$, typical of a local two-body problem.
The number of Carleman variables at this stage is thus
$bL$ for $f_i(x)$, and $b(b+1)L/2$ for $f_{ij}(x_i,x_j)$.
\begin{figure*}
\label{CarlePairs}
\centering
\includegraphics[scale=0.5]{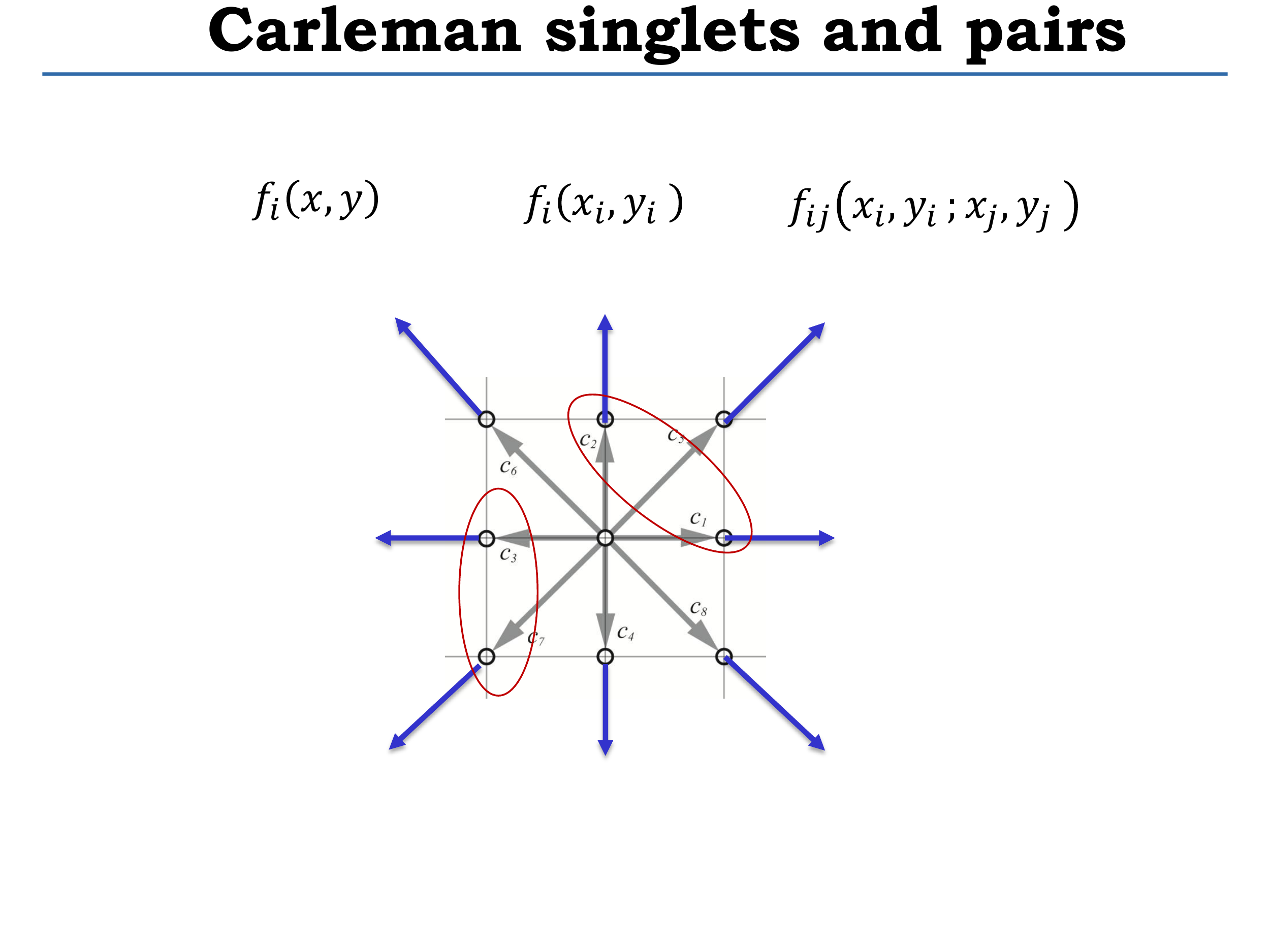}
\caption{The first three Carleman structures: local singlet $f_i(x,y)$ (central grey), 
non-local singlet $f_i(x_i,y_i)$ (blue arrows) and nonlocal pair $f_{ij}(x_i,y_i,x_j,y_j)$ (red circled).
The encircled pairs correspond to $f_{12}$ and $f_{37}$ respectively. 
The figure refers to a two-dimensional D2Q9 for better visualization purposes. 
}
\end{figure*}
\subsection{Carleman Structure for Kinetic Theory versus Fluid Dynamics} 

The above consideration signals a combinatorial many-body proliferation of Carleman 
variables, scaling like the size of the grid to some power 
associated with the Carleman truncation order, $b^2$
at order 2, $b^3$ at order 3, and so on.

In a nutshell, {\it Carleman linearization at order $k$ transforms 
a nonlinear one-body problem $d$ into a linear k-body problem.} This is the principal effect of nonlocality.
 
But let us for the moment suspend the issue of nonlocality 
and focus on local nonlinearity, namely the streaming-free 
homogeneous LB.

\subsection{Homogeneous Case}
Equation \ref{LBC1} takes the form:
\begin{equation}
\label{LBC1homo}
f_i(x,t+1) = (1-\omega) f_i(x,t) + \omega f_i^{eq}(x,t)
\end{equation}
This is a set of $b$ ODE's, whose local attractor is $f_i^{eq}$.
The equation for $f_{ij}(x,t)$ now reads simply as:
\begin{widetext}
\begin{equation}
\label{LBC2}
f_{ij}(x,t+1) 
= (1-\omega)^2 f_{ij}(x,t)  
+ 2 \omega h_{ij}(x,t) 
+ \omega^2 g_{ij}(x,t)
\end{equation}
\end{widetext}

The second order CLB reads formally as 
\begin{widetext}
\begin{eqnarray}
\label{LBC2T}
f_{i}(t+1) = [(1-\omega) \delta_{ij} + \omega L_{ij}] f_j 
+\omega  Q_{ijk} f_{jk}\\
f_{ij}(t+1) = (1-\omega)^2 f_{ij}  
+ \omega (1-\omega)[L_{jk} f_{ki} + L_{il} f_{lj}]
+ \omega^2 L_{ik} L_{jl} f_{kl}.
\end{eqnarray}
\end{widetext}

This is elegant and can be readily generalized to higher orders.
However, it shows a steep power-law growth, $N^k$ at the $k^{th}$
Carleman order, which becomes rapidly unsustainable on classical
computers.

\subsection{From Truncation to Exact Closure}
\label{subsec:truncexact}
The above treatment does not make use of the conservation laws and resulting
sum rules, relations in Eq.~(\ref{SUMRULES}), which, for the homogeneous case,
permit one to close the Carleman hierarchy exactly at the second order.

The main observation is that local equilibria depend parametrically only
on the fluid density $\rho$ and the flow field $u$.
For the case of incompressible flows, we can set $\rho=1$, so that the local
equilibria, hence the quadratic term in the Carleman procedure, depend only
on the quadratic term $u^2=\sum_{ij} f_{ij} c_i c_j$ and not on 
the single components $f_{ij}$.

For the sake of concreteness, let us report the explicit calculation
for the D1Q3 case, with the following basic parameters:
$c_0=0,c_1=-1,c_2=+1$,
$w_0=4/6,w_1=1/6,w_2=1/6$,
$Q_0=-1/3,Q_1=2/3,Q_2=2/3$. The corresponding local equilibria are
\begin{eqnarray}
f_0^{eq} = \frac{2}{3}(1-\frac{u^2}{3})\\
f_1^{eq} = \frac{1}{6}(1-3u+3u^2)\\
f_2^{eq} = \frac{1}{6}(1+3u+3u^2).
\end{eqnarray}

The equation of motion for the first Carleman level 
$F_1= [f_0,f_1,f_2]$ is
$
\frac{d f_i}{dt} = -\omega (f_i-f_i^{eq}).
$
From the expression for the local equilibria, it is clear that
the quadratic coupling is confined to the $u^2=(f_2-f_1)^2$ term;
consequently, it is sufficient to define just {\it one} second-level 
Carleman variable, $F_2 = (f_2-f_1)^2]$.
On the other hand, since $u$ is left unchanged by the collision operator,
we clearly have $du^2/dt=0$.
This shows that the four-component Carleman system, given by 
$F_{12}= [f_0,f_1,f_2;(f_2-f_1)^2]$, can be truncated
exactly to the second order \cite{itaniAnalysisCarlemanLinearization2022}.
This is an example of a simple but very effective dimensional reduction
via nonlinear mapping: instead of tracking $f_1^2,f_2^2, f_1 f_2$ separately, it is
sufficient to track the single invariant $(f_1-f_2)^2$.
The same technique is readily extended to two and three-dimensions, with three 
and six extra Carleman variables, $u_x^2,u_x u_y,u_y^2$ 
and $u_x^2,u_xu_y,u_xu_z,u_y^2,u_yu_z, u_z^2$, respectively. 

Unfortunately, this wonderful property is impaired by the streaming step, 
since $u(x)$ is no longer a dynamic invariant. 
Hence, the issue of nonlocality cannot be sidestepped; due to
the combinatorial growth of the degrees of freedom for increasing Carleman 
orders, it is clear that, on classical computers, trading nonlinearity 
for higher dimensions and non-locality is a self-inflicted exertion.

Differently restated, the CLB scheme ought to be run 
on quantum computers. 

\section{Quantum CLB Algorithm}

As stated already, the development of a quantum algorithm for classical fluid dynamics
must confront two major issues: {\it nonlinearity} and {\it non-unitarity} (dissipation).
Below, we sketch a possible strategy around both, by focusing on the
details of the dynamic steps of the quantum CLB scheme: collisions and streaming. The algorithm is laid out for the incompressible case,  but could be extended the compressible case in a straightforward manner by considering inverse bosonic operators \cite{royBosonInverseOperators1995}. 

\subsection{Strategy Overview and the Equilibrium Function}
To perform the quantum algorithm of Carleman Lattice Boltzmann, we must be able to:
\begin{enumerate}
    \item Initialize $\Vec{f}(\Vec{x},t=0)$ as well as $\Vec{u}(\Vec{x},t=0)$
    \item Perform the collision step involving for each discrete density variable $f_i$ involving $f_i$ and $\Vec{u}$
    \item Uncompute $\Vec{u}(\Vec{x},t)$ (Update the register encoding the velocity along each dimension to have a value of zero as explained in Sec.~\ref{subsubsec:velocityuncomputation})
    \item Stream the discrete density variables $f_i$
    \item Apply boundary conditions
    \item Compute $\Vec{u}(\Vec{x},t)$ at each lattice site $(\forall \Vec{x})$ (After being updated/reset to zero, each of the values of the velocities are encoded again into the respective register)
    \item Repeat steps $2$ through $6$
    \item Post-process the results
\end{enumerate}

After defining the velocity as a separate variable, in line with the findings from Sec.~\ref{subsec:truncexact}, we say that the collision step approximates:
\begin{eqnarray}
    \partial_t f_i =& -\omega(f_i-f_i^{eq}) & \forall \; i \in [0,b-1]
    \\ \partial_t u_\mu =& 0 & \forall \; \mu \in [1,d]
\end{eqnarray}

We note that the equilibrium function used in lattice Boltzmann scheme is an approximation for the Boltzmann equilibrium distribution, obtained by truncating the power series of the exponential, as
\begin{eqnarray}
    f^{eq} = \frac{\rho}{(2\pi R T)^{\frac{d}{2}}} e^{-\frac{(\Vec{c}-\Vec{u})^2}{2RT}} =& \frac{\rho}{(2\pi R T)^{\frac{d}{2}}} e^{-\frac{\vec{c}^2}{2RT}} (1+ \frac{\Vec{c}\cdot\Vec{u}}{RT}+\frac{(\Vec{c}\cdot\Vec{u})^2}{2(RT)^2}-\frac{\Vec{u}^2}{2RT}+O(\Vec{u}^3))
    \\ =& \frac{\rho}{(\frac{2}{3}\pi )^{\frac{d}{2}}c^d} e^{-\frac{3\vec{c}^2}{2c^2}} (1+ 3\frac{\Vec{c}\cdot\Vec{u}}{c^2}+9\frac{(\Vec{c}\cdot\Vec{u})^2}{2c^4}-3\frac{\Vec{u}^2}{2c^2}+O(\Vec{u}^3)).
\end{eqnarray}
where $\rho$ is the fluid density, $T$ the thermodynamic temperature, and $R$ the gas constant. We may then define
\begin{eqnarray}
\label{eq:fiequ}
    f_i^{eq} =  \frac{\rho}{(2\pi R T)^{\frac{d}{2}}} e^{-\frac{(\Vec{c}_i-\Vec{u})^2}{2RT}} =&  \frac{\rho}{(2\pi R T)^{\frac{d}{2}}} e^{-\frac{\Vec{c}_i^2}{2RT}} e^{\frac{(2\Vec{c}_i\cdot\Vec{u}-\Vec{u}^2)}{2RT}}
    \\ =& \frac{\rho}{(2\pi R T)^{\frac{d}{2}}} e^{-\frac{\Vec{c}_i^2}{2RT}} e^{\frac{\Sigma_{\mu=1}^d(2c_{i,\mu}u_\mu-{u}_\mu^2)}{2RT}}
    \\ =& \frac{\rho}{(2\pi R T)^{\frac{d}{2}}} e^{-\frac{\Vec{c}_i^2}{2RT}} \Pi_{\mu=1}^d \Sigma_{k=0}^\infty \frac{1}{k!} (\frac{c_{i,\mu}}{\sqrt{2RT}})^k H_k(\frac{u_\mu}{\sqrt{2RT}}),
\end{eqnarray}
where we have identified the generating function for the Hermite polynomials,
\begin{equation}
    e^{2xy-y^2} = \Sigma_{n=0}^\infty H_n(x) \frac{y^n}{n!},
\end{equation}
where $H_n(x)$ is the $n^{th}$ Hermite polynomial, with $x = \frac{u_\mu}{\sqrt{2RT}}$ and $y = \frac{c_{i,\mu}}{2RT}$.

\subsection{Ideal Scaling}
The quantum state may be described by $\ket{\Psi(x)} \equiv \Sigma_{x}\Sigma_{c}^{C_n} \alpha_{x,c}\ket{x}\ket{c}$ with $\Sigma_{x,c}|\alpha_{x,c}|^2  = 1$ for an appropriate normalization. 
More precisely, denoting by $C_n$ the number of Carleman components at each
lattice site at the truncation level $n$, and $N=L^d$ the number 
of spatial lattice sites in $d$ spatial dimensions, the classical system can be encoded
within a set of $Q=d\log_2{L}+\log_2{C_n} = \log_2{N}+\log_2{C_n} = \log_2{NC_n}$ qubits, 
$\ket{\Psi} = \bigotimes_{q}^Q (\alpha_{q} \ket{0} + \beta_q \ket{1})$, 
where $a_q \equiv |\alpha_q|^2$ is the probability of finding the $q^{th}$ qubit in state $\ket{0}$,
and $b_q \equiv |\beta_q|^2 = 1-a_q$ is the probability of finding the same qubit in state $\ket{1}$. For the case of the lattice Boltzmann method, while accounting for the fact that the system is exactly linear with the introduction of a finite number of variable, it is sufficient to use $Q = \lceil\log_2{(b+\frac{3!}{(3-(d-1))!} )N}\rceil$ to represent the state of the system. This is the ideal scaling of the qubit complexity $Q$ of the algorithm against which we must compare the one we are able to achieve with our mapping.
\subsection{Mapping}
We encode our variables as eigenvalues of coherent states with bosonic lowering operator as eigenvector \cite{kowalskiNonlinearDynamicalSystems1997}:
\begin{equation}
    \hat{a}_i \ket{\Vec{f}(\Vec{x},t)} = f_i \ket{\Vec{f}(\Vec{x},t)}  \forall \; i \in [0,b-1]
\end{equation}
and 
\begin{equation}
    \hat{a}_\mu \ket{\Vec{u}(\Vec{x},t)} = u_\mu \ket{\Vec{u}(\Vec{x},t)}  \forall \; \mu \in [1,d]
\end{equation}
where
\begin{equation}
    \ket{\Vec{f}(\vec{x},t)} = \bigotimes_{i=0}^{b-1} \ket{f_i(\Vec{x},t)}
\end{equation}
and
\begin{equation}
    \ket{f_i(\Vec{x},t)} = \Sigma_{m=0}^\infty \frac{1}{m!}f_i^m(\Vec{x},t) \ket{m}
\end{equation}
in the occupation number basis, with the set of statevectors $\ket{m}$ representing its eigenvectors, up to an appropriate normalization. Of course, for implementation on a quantum computer, the number $n$ of excitation levels considered is finite, and the implications of this observation are discussed in \cite{itaniBinaryRepresentationTruncated2022}. It is straightforward to see how coupled terms arise from the tensorial coherent states. For example,
\begin{equation}
    \ket{f_i}\ket{f_j} = (\Sigma_{k=0}^\infty \frac{1}{k!}f_i^k(\Vec{x},t) \ket{k})(\Sigma_{m=0}^\infty \frac{1}{m!}f_j^m(\Vec{x},t) \ket{m}) = \Sigma_{k,m}^\infty \frac{1}{k!m!}f_i^k (\Vec{x},t)kf_j^m(\Vec{x},t) \ket{k,m}.
\end{equation}
Rather than considering higher excitation levels, we suffice with one excitation level upon the introduction of the second-order variables, as
\begin{eqnarray}
    \ket{f_i^2} & \forall \; i \in [0,b-1]
    \\ \ket{u_\mu^2} & \forall \; \mu \in [1,d].
\end{eqnarray}
Here and elsewhere, we use
\begin{eqnarray}
    \ket{f_{i,1}} =& \ket{f_i}
    \\\ket{f_{i,2}} =& \ket{f_i^2},
\end{eqnarray}
 and similarly for the case of the registers encoding the velocity. The corresponding bosonic operators are labeled as $\hat{a}_{(f,u),(i,\mu),m}$, e.g., $\hat{a}_{f,1,2}$ corresponding to $\ket{f_i^2}$. Moreover, we define
 
 \begin{eqnarray}
 \Omega_{i,1} =& \Omega_i
 \\\Omega_{i,2} =& 2f_i\Omega_i
 \end{eqnarray}
 where
 \begin{eqnarray}
     \Omega_i(\Vec{f}) = -\frac{1}{\tau}(f_i(\Vec{x},t)-f_i^{eq}(\Vec{x},t))
 \end{eqnarray}
 not to be confused with $\Omega = \frac{1}{\tau}$ which defines the relaxation frequency for $\Omega_i$.
 Since all bosonic Fock space only includes a single excitation level, a number of qubits equals the number of extended variables $(2(b+d))$, along with the $\log_2{L^d}+\log_2{d}$ term required for streaming \cite{itaniQuantumAlgorithmLattice2022}, making the total number of variables $2b+d(\log_2{L}+2)+\log_2{d}$.

\subsection{Initialization and Evolution}
The discrete density and the velocity registers are initialized using the corresponding unitary displacement operators
\begin{equation}
    \hat{D}(\alpha) \ket{0} = e^{\alpha \hat{a}^\dagger-\alpha^{*}\hat{a}} \ket{0}= \ket{\alpha}.
\end{equation}

Next, we write the evolution equation for the 
quantum state:
\begin{equation}
    \ket{\Psi(t)} = \Sigma_{\vec{x}}\bigotimes_{m=1}^2\bigotimes_{i=0}^{b-1}\bigotimes_{\mu=1}^{d} \ket{\vec{x}} \ket{f_{i,m}} \ket{u_{\mu,m}}
\end{equation} 

\begin{equation}
\partial_t \ket{\Psi} =(-\hat{S}+\hat{C}) \ket{\Psi},
\end{equation} 
where $\hat{S}=c \partial_x$ is the streaming operator and $\hat{C}$ is 
the Carleman-linearized collision operator. 

The time-propagator over a time-step $\Delta t$ reads formally as
$
\hat{T}_{\Delta t} = e^{(-\hat{S}+\hat{C})\Delta t}.
$
Given that streaming and collision do not commute, the above
propagator must be treated via standard Trotter factorization. Luckily, the lattice Boltzmann method assumes that streaming and collision occur in separate steps, and any Trotterization of the unseparated propagator can be shown to be equivalent to using a finer lattice (larger number of lattice sites). The task of the quantum algorithm is to devise a 1:1 map between
the separated time-propagator and a suitable quantum circuit.

\subsubsection{Collision Operator}
Since it is a local operator it acts only on the Fock space modes, and not on the lattice position register, benefiting from quantum parallelism. Collisions are responsible for relaxation to local equilibrium, hence they break 
reversibility and introduce dissipation, which means that the collisional
propagator 
\begin{equation}
\hat{T}^C_{\Delta t} = e^{\Delta t \hat{C}}
\end{equation}
is non-unitary. A collision that changes, say, $f_1(x=3)$ into $f'_1(x=3)$ corresponds to a change
of parameters $b_1 a_2 a_3=f_1(x=3)$ to $b'_1 a'_2 a'_3=f'_1(x=3)$.  This is a rotation
in the Block spheres of the three qubits, followed by a contraction, and can
be implemented as the weighted sum of two unitaries.
This mapping extends to higher dimensions as well.

Explicitly, $\hat{C}$ can be written as:
\begin{eqnarray}
    \hat{C} =& \Sigma_{m=1}^2\Sigma_{i=0}^{b-1} \hat{a}^\dagger_{f,i,m}\Omega_{i,m}(\hat{a}_{f,i,1},\hat{a}_{f,i,2},\hat{\vec{a}}_{u,1},\hat{\vec{a}}_{u,2})
    \\ =&\Sigma_{m=1}^2\Sigma_{i=0}^{b-1} \hat{a}^\dagger_{f,i,m} \hat{a}_{f,i,m-1}\omega(\hat{a}_{f,i,m}-\frac{1}{(2\pi R T)^{\frac{d}{2}}} \Pi_{\mu=1}^d e^{-\frac{c_{i,\mu}^2-2c_{i,\mu}\hat{a}_{u,\mu,1}+\hat{a}_{u,\mu,2}}{2RT}})
\end{eqnarray}
where $\vec{\hat{a}}_{u,1} = (\hat{a}_{u,1,1},\dots,\hat{a}_{u,\mu,1},\dots,\hat{a}_{u,d,1})$, since we encoded the variables as eigenvalues of coherent states \cite{kowalskiNonlinearDynamicalSystems1997}. Then the operator $\hat{C}$ weighted by $\Delta t$ achieves a displacement of $\ket{\Vec{f}(\Vec{x},t)}$ by $\Delta t\vec{\Omega}(\Vec{f})$ when exponentiated. In that sense, collision is achieved by a term $\Vec{a}^\dagger\cdot \Delta \vec{\Omega}{\Vec{(f)}} = \Vec{a}^\dagger\cdot \Delta \vec{\Omega}{\Vec{(\hat{a})}}$ similar to one appearing in the exponential of a displacement operator $\hat{a}^\dagger\alpha$ which is antisymmetrized with $-\alpha^*\hat{a}$ to achieve a unitary operator ($\hat{-i\Delta t H}$ is anti-Hermitian for a Hermitian $\hat{H}$). The Hermitization of the ladder operators corresponds to the position-momentum operators.

A few notes are due. First, when considering the practical implementation of the raising and lowering operators, their truncation to $n$ levels means that if the bosonic particles are raised or lowered beyond $n$ times, they are no longer accounted for inside the system. If the highest occupied level is the $n^{th}$ level, then applying the lowering operator $n$ times annihilate all other particles initially occupying levels lower than $n$, and brings the particles in the $n^{th}$ excitation level to the $0^{th}$ level. As such, the application of the lowering operator $n+1$ times guarantees the annihilation of all particles in the system. Similarly, if the the ground level is the lowest level, the application of the raising operator $n$ times moves all particles not initially in the ground level to an excitation level beyond the truncation at level $n$, and the particles initially in the ground state to the $n^{th}$ excitation level. As such, the application of the raising operator $n+1$ times guarantees the annihilation of all particles in the system which would have moved to levels higher than $n$ were the bosonic Fock space not truncated. Mathematically, this is seen in a more straighforward manner by noting that
\begin{eqnarray}
(\hat{a})^{n+1} = (\hat{a}^\dagger)^{n+1} = 0
\end{eqnarray}
for the truncated bosonic operators $\hat{a}$, and $\hat{a}^\dagger$:
\begin{equation}
    \hat{a} = (\hat{a}^\dagger)^\dagger = \begin{pmatrix}
        0 & 0 & \dots & 0 & 0
        \\\sqrt{1} & 0 & \dots & 0 & 0
        \\0 & \sqrt{2} & \ddots & \vdots & \vdots
        \\ 0 & 0 & \ddots & 0 & 0
        \\ 0 & 0 & \dots & \sqrt{n} & 0
    \end{pmatrix}
\end{equation}
Thus, we say that $\hat{a}$ and $\hat{a}^\dagger$ are nilpotent to the power $n+1$ i.e. their $(n+1)^{th}$ power is zero. Therefore, truncating the bosonic Fock space guarantees the truncation of the equilibrium distribution since
\begin{eqnarray}
    e^{-\frac{c_{i,\mu}^2-2c_{i,\mu}\hat{a}_{u,\mu,1}+\hat{a}_{u,\mu,2}}{2RT}}
    \\=& e^{-\frac{c_{i,\mu}^2}{2RT}}e^{\frac{c_{i,\mu}}{2RT}\hat{a}_{u,\mu,1}}e^{-\frac{\hat{a}_{u,\mu,2}}{2RT}} 
    \\=& e^{-\frac{c_{i,\mu}^2}{2RT}} (\Sigma_{k=0}^\infty \frac{1}{k!} (\frac{c_{i,\mu}}{RT})^k(\hat{a}_{u,\mu,1})^k)(\Sigma_{k'=0}^\infty \frac{1}{(k')!} (\frac{-1}{RT})^{k'}(\hat{a}_{u,\mu,2})^{k'})
    \\=& e^{-\frac{c_{i,\mu}^2}{2RT}} (\Sigma_{k=0}^n \frac{1}{k!} (\frac{c_{i,\mu}}{RT})^k(\hat{a}_{u,\mu,1})^k)(\Sigma_{k'=0}^n \frac{1}{(k')!} (\frac{-1}{RT})^{k'}(\hat{a}_{u,\mu,2})^{k'}),
\end{eqnarray}
where the last equality holds due to the truncation.

In fact, since the system is relaxing to a local equilibrium, its eigenvalues must be
real and positive, hence $\hat{T}_C$ is a contraction. The corresponding quantum gate is therefore in charge of transforming the 
parameters of the qubit configuration according to
the change from $f_i$ to $f'_i$. 

\begin{figure*}
\label{Qcolli}
\centering
\includegraphics[scale=0.3,angle=-00]{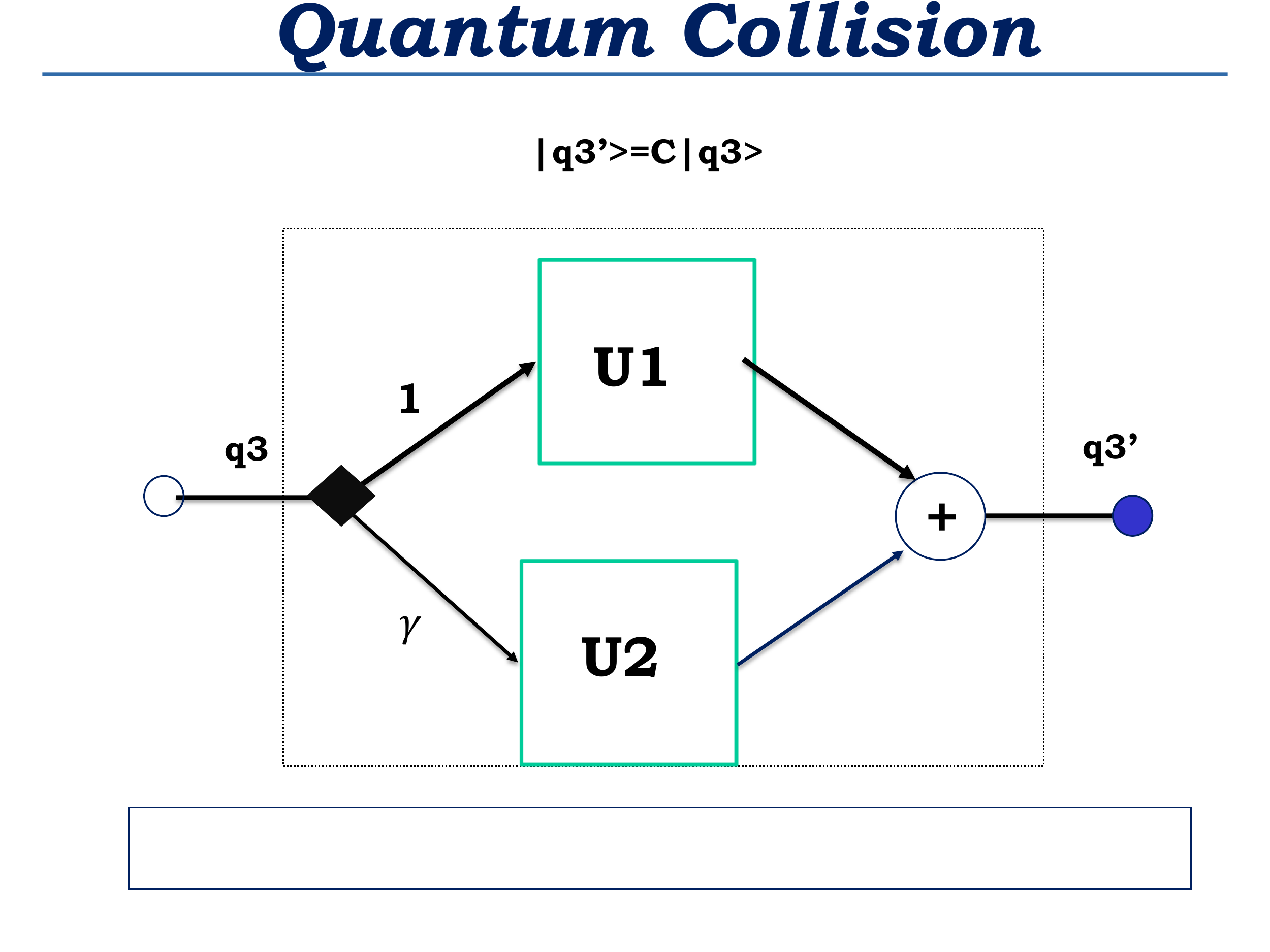}
\caption{Quantum circuit implementing the collision operator as a linear combination of unitaries.
The gates $U1$ and $U2$ are executed with probability ratio $1:\gamma$,
to implement a statistically averaged weighted sum.
}
\end{figure*}
 
A possible way to deal with non-unitary transfer 
operators is to represent them as a weighted sum of unitaries \cite{childsHamiltonianSimulationUsing} as in
$
\hat{T}^C_h = \hat{U}_1 + \gamma \hat{U}_2,
$
where $\gamma$ is a real-valued scalar. The scalar $\gamma$ can be chosen to minimize 
the probability of failure. It worthy to note the possibility of iterative application of the procedure for the case where the linear combination of unitaries involves more than two unitaries. The procedure has been successfully implemented for the case of linear
advection-diffusion operators \cite{mezzacapoQuantumSimulatorTransport2015,budinskiQuantumAlgorithmAdvection2021}, but the nonlinear case remains to be explored. 

\subsubsection{Velocity Uncomputation}
\label{subsubsec:velocityuncomputation}
Once the collision operation is performed, the values encoded into the registers representing the velocities across each dimension need to be reset back to zero before the streaming operator is applied. The concept behind this is that the expressions for the square of each the velocities contain cross terms, e.g. in D1Q3 $u^2 = (f_1-f_3)^2 = f_1^2-2f_1f_3+f_3^2$ which can not be streamed exactly, but only through the encoding of position of lattice sites into an occupation number basis like other variables, as suggested in Sec.~\ref{eq:altstreaming} rather than the simple binary representation employed. 

Consequently, the velocities are reset to zero, uniformly over the lattice, such that no streaming is needed, and reevaluated (recalculated) at each lattice site after the streaming of discrete density variables. 

\begin{itemize}
    \item Uncomputation: $\Vec{u}(\Vec{x},t) \xrightarrow{-\Vec{u}(\Vec{x},t)}\vec{0}$ or $\Vec{u}(\Vec{x},t) \xrightarrow{-\Vec{u}(\Vec{f})}\Vec{0}$
    \item Computation: $\vec{0}\xrightarrow{+\Vec{u}(\Vec{f})} \Vec{u}(\Vec{x},t)$
\end{itemize}

This recalculation, of course, happens in parallel similar to the initialization procedure. In order to reset the velocities to zero, we may either use the bosonic operators associated with discrete densities appearing in its expression, or those with the velocity variable itself, to displace the coherent state encoding the velocity. In D1Q3, for example, we apply the operator
\begin{equation}
    e^{-\hat{a}^\dagger_{u,1,2}(\hat{a}_{f,1,2}-2\hat{a}_{f,1,1}\hat{a}_{f,3,1}+\hat{a}_{f,3,2})-\hat{a}^\dagger_{u,1,1}(\hat{a}_{f,1,1}-\hat{a}_{f,3,1})}.
\end{equation}

In practical terms, this process resets the velocities back to zero up to an error which scaling is best 
discussed separately in the context of the implementation of the algorithm we propose here. Moreover, the operators appearing as a linear combination do not necessarily commute.

Alternatively,  given that the velocity is conserved under collision,  we use the expression of the velocity in terms of the discrete densities, whose 
value is conserved despite the discrete density variables being updated. Again, in D1Q3, this would correspond to the operator
\begin{equation}
    e^{-\hat{a}^\dagger_{u,1,1}\hat{a}_{u,1,1}-\hat{a}^\dagger_{u,1,2}\hat{a}_{u,1,2}} = e^{-\hat{n}_{u,1,1}}e^{-\hat{n}_{u,1,2}}.
\end{equation}

In both cases, we see that the register for each power of the velocity variable, $u$ and $u^2$, is acted upon. In the second case, where there is only need to act on the corresponding register for each of the powers, we see that the operator becomes separable. The expressions for either approach are readily extensible to higher dimensions. 

\subsubsection{Streaming Operator}

Streaming can be mapped into quantum gates by casting 
the kinetic equation in the language of second quantization in
imaginary time. The second-quantized streaming transfer operator takes the form
$
\hat{T}^S_{\Delta t} = e^{c\Delta t(\hat{a}^\dagger-\hat{a})}, 
\label{eq:altstreaming}$
where $\hat{a}$ and $\hat{a}^+$ are the standard generation-annihilation 
operators and $\Delta t$ is the time step.
This streaming operator is unitary, which could readily be seen by the fact that the streaming operator coincides with a displacement of $\ket{\Vec{x}}$ defined to be a coherent state. Its expression involves operator powers at all orders, each order corresponding to
the excited state of a corresponding pseudo-spin quantum bosonic system, which
can be physically realized, for instance, by ion traps \cite{mezzacapoQuantumSimulatorTransport2015}.
The fact that the exact propagator contains excitations at
all orders implies a truncation error in the Fock space of the bosonic 
system because no physical system can support an infinite tower of bosonic excitations. Details of this mapping, namely the identification of the quantum circuits
associated with the streaming operator, are described in earlier publications \cite{succiLatticeBoltzmannEquation1993,mezzacapoQuantumSimulatorTransport2015,itaniQuantumAlgorithmLattice2022}. 

Instead, an exact streaming procedure could be achieved by defining $\ket{x_j}$ to encode binary representation of lattice index along the $j^{th}$ dimension, following the exact streaming procedure described in \cite{todorovaQuantumAlgorithmCollisionless2020a,itaniQuantumAlgorithmLattice2022}.

\begin{figure*}
\centering
\includegraphics[scale=0.3,angle=-00]{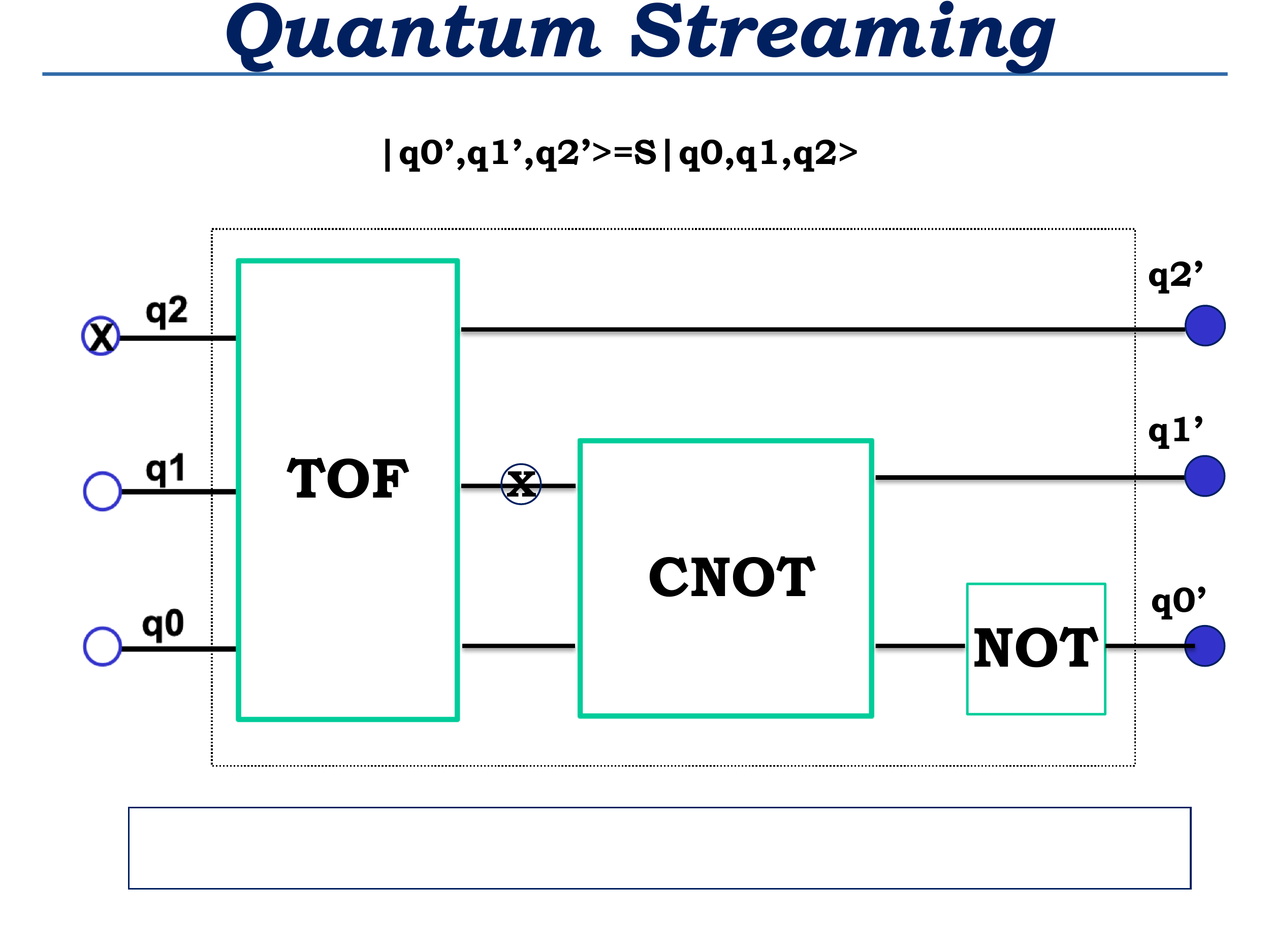}
\caption{\label{fig:Qstream} Quantum circuit implementing the right-streaming operator
with periodic boundary conditions.
Crossed circles indicates target qubits.}
\end{figure*}

\begin{figure*}
\centering
\includegraphics[scale=0.3]{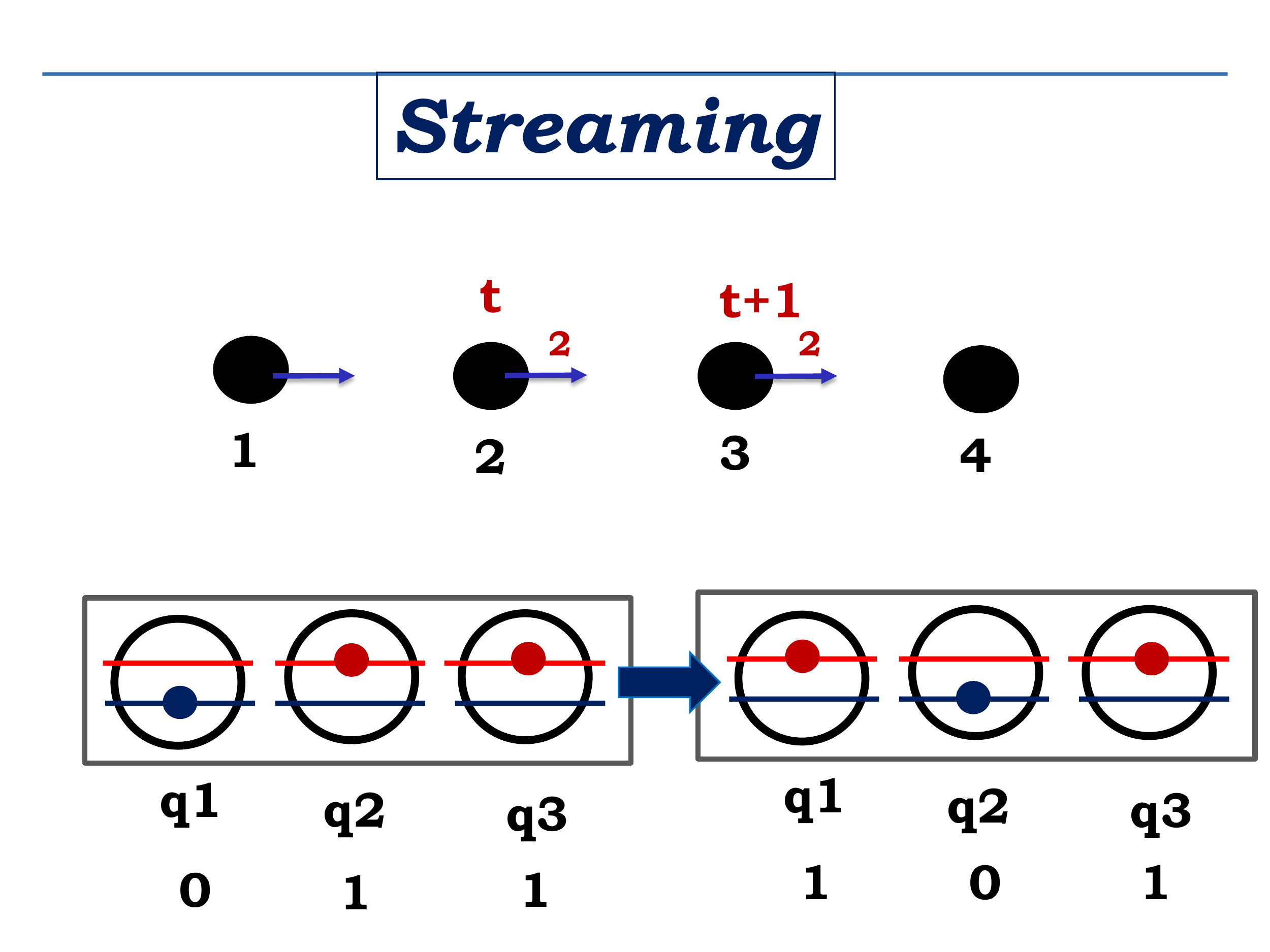}
\caption{Quantum map of the streaming operator.
$\ket{011} \to \ket{101}$ codes the streaming $f_1(3,t+1)=f_1(2,t)$.
}
\end{figure*}
\subsubsection{Velocity Computation}
The streaming operator must be followed by computation of the velocity $\Vec{u}(\Vec{x},t)$. This is done by taking the inverse operation of the computation discussed above, which also benefits from quantum parallelism similar to the collision operator.

\section{Tentative Scaling of Quantum Simulations}

Current leading edge fluid simulations run on about 10 trillion 
grid points, $N=10^{13}$, corresponding to about $Re \sim 10^5$ -- $10^6$. 
As a result, we can consider the quantum advantage at $N=10^{15}$, corresponding 
to $Re \sim 10^7$.  

Major barriers stand in the way of such a prospective quantum advantage.
Leaving aside the notorious problem of noise, which is common to any quantum implementation,
there are specific issues related to the quantum CLB algorithm, independent
of its actual quantum implementation, particularly the
convergence and scalability of the Carleman linearization as a function 
of the Reynolds number. 
 
The convergence and scalability of the Carleman procedure for systems of quadratic ODEs 
has been studied in detail \cite{liuEfficientQuantumAlgorithm2021} with the use of the QLAS quantum linear 
algebraic solver \cite{harrowQuantumAlgorithmLinear2009a}. 
The main result is that the complexity of the algorithm scales like
$ \mathcal{C} \sim s T^2 F Poly(log N, logT,log1/\epsilon)$, where 
$N$ is the size of the Carleman-linearized system and $s$ its sparsity,
$T$ the time lapse of the simulation, $F$ the overlap of the initial state 
with the final one (fidelity) and $\epsilon$ the error tolerance.  
This is an important result, as it circumvents the exponential barrier in 
$T$, of previous formulations.
However, it only holds for $Re<1$, which rules out the vast majority 
of macroscopic flows, let alone turbulence. With the numbering convention described above, the Carleman matrix 
shows a block-sparse structure, with full blocks of size $C_n \times C_n$
for the collision matrix and $C_n$ sparse entries for the streaming operator.
It would be interesting to study whether such a structure lends itself to
general QLAS techniques.

Quantum simulation of the Burgers equation
indicates a notably higher threshold, $Re \sim 40$, showing 
that the theory is over-restrictive, a point which begs for
further investigations, possibly in the direction of unanticipated
error cancellation rather than accumulation.
At this stage it is worth emphasizing that the quantum algorithm proposed
in this paper differs considerably from that in \cite{liuEfficientQuantumAlgorithm2021,anEfficientQuantumAlgorithm2022}. They both
rely on Carleman linearization and 
resort to a QLAS, 
whereas our proposal here is a direct mapping onto
a quantum analogue physical device. 
Being tailored to the (lattice kinetic) theory of fluids, our approach 
is less general but possibly more efficient.
Most importantly, the Navier-Stokes case is vastly more complex than Burgers: it is three-dimensional and, more importantly, requires the tracking
of pressure-velocity-stress correlations.
As discussed in the early part of this paper, 
we expect the kinetic formalism presented here
to offer significant simplifications, although this can only 
be proven by actual quantum simulations. 
In particular, the limiting Reynolds number for the Carleman procedure 
discussed in this paper is unknown and stands as the key question to 
decide about the practical viability of quantum computing for fluid turbulence.

\section{Summary}

We have presented a prospective quantum computing algorithm 
for the solution of classical fluid dynamics based on the Carleman 
linearization of the Lattice Boltzmann equation.
The main result is that the CLB formulation largely preserves
the structure of the LB algorithm, although the streaming step
entails a growth of nonlocality, which is hardly handled
by a classical algorithm for all but the lowest Carleman levels. 

At variance with previous formulations, our algorithm does not rely on 
quantum linear algebra for the solution of Carleman-linearized system,
but rather on a second-quantized formulation of the kinetic equation
which maps directly onto a quantum physical device, hence
custom-made for the lattice kinetic formulation of fluid dynamics.

The details of the quantum circuit, hence its quantum scalability,
remain to be tightened and its practicality assessed by actual 
simulations on quantum hardware.

The quantum advantage for high-Reynolds-number flows remains entirely open at this stage. On philosophical grounds, it would not be surprising if high Reynolds
numbers would prove too hard for quantum computing. 
Indeed, coming back to Feynman, while it is true that Nature is not classical, it is
equally true that Nature has a very strong innate tendency to {\it become}
classical at macroscopic scales. In this respect, the prominence of non-locality
in exchange for the release of nonlinearity, may well represent yet another 
signature of this tendency.

Figuring this out is a worthy enterprise, regardless of the practical 
outcome.  Moreover, we wish to observe that
the physics of fluids is populated with interesting problems
at low Reynolds numbers, especially in soft matter and biological flows \cite{bernaschiMesoscopicSimulationsPhysicschemistrybiology2019}.
For instance, it would be of great interest to devise a
{\it quantum multiscale} application, coupling quantum algorithms for 
biomolecules swimming in a water solvent described by a quantum 
algorithm for low-Reynolds-number flow.   

\begin{acknowledgments}
One of the authors (SS) is grateful to the SISSA program on 
"Collaborations of Excellence" that allowed him to visit and focus on the 
work presented in this paper. 
Illuminating discussions with S. Ruffo and A. Solfanelli are kindly acknowledged. 
He also wishes to acknowledge support from National Centre for HPC, Big Data 
and  Quantum Computing” (Spoke 10, CN00000013).

\end{acknowledgments}

\bibliography{references}
\end{document}